\documentclass[10pt]{article}
\def\paradot#1{\vspace{1.3ex plus 0.7ex minus 0.5ex}\noindent{\bf\boldmath{#1.}}}
\usepackage{natbib}
\usepackage{latexsym,amsmath,amssymb,amsthm,amsfonts,graphicx}
\include{def}
\def\,{\mskip 3mu} \def\>{\mskip 4mu plus 2mu minus 4mu} \def\;{\mskip 5mu plus 5mu} \def\!{\mskip-3mu}
\def\dispmuskip{\thinmuskip= 3mu plus 0mu minus 2mu \medmuskip=  4mu plus 2mu minus 2mu \thickmuskip=5mu plus 5mu minus 2mu}
\def\textmuskip{\thinmuskip= 0mu                    \medmuskip=  1mu plus 1mu minus 1mu \thickmuskip=2mu plus 3mu minus 1mu}
\textmuskip
\def\be{\dispmuskip\begin{equation}}    \def\ee{\end{equation}\textmuskip}
\def\beqn{\dispmuskip\begin{displaymath}}\def\eeqn{\end{displaymath}\textmuskip}
\def\bea{\dispmuskip\begin{eqnarray}}    \def\eea{\end{eqnarray}\textmuskip}
\def\bqan{\dispmuskip\begin{eqnarray*}}  \def\eqan{\end{eqnarray*}\textmuskip}

\def\bqan{\dispmuskip\begin{eqnarray*}}  \def\eqan{\end{eqnarray*}\textmuskip}

\def\v{}

\def\I{1\!\!1} 				

\def\g{\gamma}
\def\S{\Sigma}
\def\s{\sigma}

\def\b{\beta}
\def\l{\lambda}

\def\MLE{\text{\rm MLE}}

\renewcommand{\baselinestretch}{1}
\title{Bayesian Adaptive Lasso}
\author{Chenlei Leng, Minh Ngoc Tran and David Nott
  \footnote{Leng, Tran and Nott are with 
    Department of Statistics and Applied Probability, 
    National University of Singapore. 
    Corresponding author: Chenlei Leng,
    Email: stalc@nus.edu.sg, 
    Phone: 65-65164462. } 
 }
\date{\today}
\begin{document}
\maketitle
\begin{abstract}
  We propose the Bayesian adaptive Lasso (BaLasso) for variable selection and
  coefficient estimation in linear regression. The BaLasso is adaptive to
  the signal level by adopting different shrinkage for different coefficients.
  Furthermore, we provide a model selection machinery for the BaLasso by 
  assessing the posterior conditional mode estimates, motivated by the
  hierarchical Bayesian interpretation of the Lasso. Our formulation also 
  permits prediction using a model averaging strategy. We discuss other variants
  of this new approach and provide a unified framework for 
  variable selection using flexible penalties. 
  Empirical evidence of the attractiveness of the method is demonstrated via
  extensive simulation studies and data analysis.
\end{abstract}
  
\section{Introduction}
Consider the linear regression problem
\[ y =\mu 1_n +X\beta+\epsilon,
\] 
where $y$ is an $n \times 1$ vector of responses, $X$ is an $n \times p$ matrix of 
covariates and $\epsilon$ is an $n \times 1$ vector of iid normal errors
with mean zero and variance $\sigma^2$. As is usual in regression analysis, our major interests are to estimate $\beta=(\beta_1,...,\beta_p)'$, to identify
its important covariates and to make accurate predictions. 
Without loss of generality, we assume $y$ and $X$ are centered
so that $\mu$ is zero and can be omitted from the model.

In an important paper, Tibshirani (1996) proposed the least absolute shrinkage and
selection operator (Lasso) for simultaneous variable selection and 
parameter estimation. The Lasso, formulated in the penalized likelihood 
framework, minimizes the residual sum of squares with a constraint on 
the $\ell_1$ norm of $\beta$. Formally, the Lasso solves
\be
\underset{\beta}{\mbox{min}} ~(y-X\beta)^T(y-X\beta)+\lambda \sum_{j=1}^p |\beta_j|,
\label{eq1}
\ee
where $\lambda >0$ is the tuning parameter controlling the amount of penalty. 
The least angle regression (LARS) algorithm provides fast implementation
of the Lasso solution \citep{Efron:2004,Osborne:2000}. 
Furthermore, the Lasso can be model selection consistent provided that the so-called irrepresentable condition on the design matrix is satisfied and that $\lambda$ is chosen judiciously \citep{Zhao:2006}.

However, if this condition does not hold, Zou (2006) and Zhao and Yu (2006)
showed that the Lasso chooses the wrong model with non-vanishing probability, regardless of the sample size and how $\lambda$ is chosen. The condition is almost necessary and sufficient for model selection consistency of Lasso, which requires that the predictors not in the model are not representable by predictors in the true model. This condition can be easily violated due to the collinearity between the predictors. To address this issue, Zou (2006) and Wang et al. (2006) proposed to use adaptive Lasso (aLasso) which gives consistent model selection. The final inference procedure, thereafter, is based on a single selected model. This may bring undesirable risk properties as discussed by 
\cite{Poetscher:2009}. \cite{Meinshausen:2009} introduced sub-sampling in model selection that improves the Lasso.

The Lasso estimator can be interpreted as the posterior mode using
normal likelihood and iid Laplace prior for $\beta$ (Tibshirani, 1996). 
Yuan and Lin (2006) studied an empirical Bayes variable selection method targeting at finding this mode. 
The first explicit treatment of the Bayesian Lasso (BLasso), which exploits model 
inference via posterior distributions, has been proposed by Park and Casella (2008). 
Hans (2010) considers a formal Bayesian approach to exploring model uncertainty with lasso type priors on parameters in
submodels.  Griffin and Brown (2010) have previously considered generalizing the Bayesian 
lasso in various ways including the use of  
separate scale parameters for different coefficients in the Laplace prior 
with gamma mixing distributions for the scale parameters.  
This is similar to the priors we use here, but Griffin and Brown (2010) focused on finding posterior mode 
estimates via an EM algorithm whereas our objectives here are somewhat broader.  In particular we aim to investigate
MCMC computational methods for these priors, estimates of regression coefficients other than the mode, 
different choices for smoothing parameters, model averaging strategies which explore model uncertainty for
predictive purposes and generalizations beyond the linear model.  

Although the Lasso was originally designed for variable selection, the BLasso 
loses this attractive property, not setting any of the coefficients to
zero. A post hoc thresholding rule may overcome this difficulty but it brings 
the problem of threshold selection. Alternatively, 
Kyung et al. (2009) recommended 
to use the credible interval on the posterior mean. 
Although it gives variable selection, this suggestion fails to explore the
uncertainty in the model space. 
On the other hand, the so-called spike and slab prior, in which the
scale parameter for a coefficient is a mixture of a point mass at zero and
a proper density function such as normal or double exponential (Yuan and Lin, 2005), 
allows exploration of model space at the expense of increased computation for a full Bayesian posterior. 

This work is motivated by the need to explore model uncertainty and to 
achieve parsimony. With these objectives, we consider the following adaptive Lasso 
estimator:
\be
\underset{\beta}{\mbox{min}} ~(y-X\beta)^T(y-X\beta)+ \sum_{j=1}^p 
\lambda_j|\beta_j|,
\label{eq2.1}
\ee
where different penalty parameters are used for the regression coefficients.
Naturally, for the unimportant covariates, we should put larger penalty parameters $\lambda_j$ on their corresponding coefficients. This strategy was proposed
by Zou (2006) and  Wang et al. (2006) by using some preliminary
estimates of $\beta$ such as the least-squares estimate $\hat\beta^0$ and modifying 
$\lambda_j$ as $\lambda/|\hat\beta_j^0|$. 
Our treatment is completely different and is motivated by the
following arguments. Suppose tentatively that we have a posterior distribution on $\{\lambda_j\}_{j=1}^p$. By drawing random samples from this distribution and plugging these into (\ref{eq2.1}), we can solve for $\beta$ using fast algorithms developed for Lasso (Efron et al., 2004; Figueiredo et al., 2007) and subsequently obtain an array of (sparse) models. These models can be used not only for exploring model uncertainty, but also for prediction with a variety of methods akin to Bayesian model averaging. Since there are $p$ tuning parameters, a hierarchical model is proposed to alleviate the problem of estimating many parameters. We develop an efficient Gibbs sampler for posterior inference.

The BaLasso permits a unified treatment for variable selection with flexible penalties, using the least sqaures approximation (Wang and leng, 2007).
 The extension encompasses generalized linear models, Cox's model and other parametric models as special cases. We outline novel applications of BaLasso when structured penalties are present, for example, grouped variable selection (Yuan and Lin, 2007) and variable selection with a prior hierarchical structure (Zhao, Rocha and Yu, 2009).

The rest of the paper is organized as follows.
The Bayesian adaptive Lasso (BaLasso) method is presented in Section \ref{secBaLasso}. Furthermore, we propose two approaches for estimating the tuning parameter vector $\lambda=(\lambda_1,...,\lambda_p)'$
and give an explanation for the shrinkage adaptivity. 
Section \ref{secInf} discusses model selection and Bayesian model averaging.
In Section \ref{secExam}, the finite sample performance of BaLasso is 
illustrated via simulation studies, and analysis of two real datasets. Section 5 presents a unified framework which deals with variable selection in models with structured penalties. Section 6 gives concluding remarks. A Matlab implementation is available from the authors' homepage.
The software is very general and deals with many parametric models encountered in practice.

\section{Bayesian Adaptive Lasso}\label{secBaLasso}

The $\ell_1$ penalty corresponds to a conditional Laplace prior (Tibshirani, 1996) as
\[\pi(\beta|\sigma^2) = \prod_{j=1}^p \frac{\lambda}{2\sqrt{\sigma^2}} e^{-\lambda|\beta_j|/\sqrt{\sigma^2}},
 \]
which can be represented as a scale mixture of normals with an exponential
mixing density (Andrews and Mallows, 1974)
\[ \frac{\lambda}{2} e^{-\lambda|z|}
=\int_0^\infty \frac{1}{\sqrt{2\pi s}}e^{-z^2/(2s)}\frac{\lambda^2}{2}
e^{-\lambda^2z/2}ds.
\] 
This motivates the following hierarchical BLasso model 
\citep{Park:2008}
\bea\label{eq2}
y|X,\beta,\sigma^2 &\sim & N_n(X\beta,\sigma^2 I_n)\notag\\
\beta|\sigma^2,\tau_1^2,...,\tau_p^2&\sim&N_p(0_p,\sigma^2 D_\tau)\\
D_\tau&=&\mbox{diag}(\tau_1^2,...,\tau_p^2)\notag
\eea
with the following priors on $\sigma^2$ and $\tau=(\tau_1^2,...,\tau_p^2)$:
\be\label{eq3} 
\sigma^2,\tau_1^2,...,\tau_p^2\sim\pi(\sigma^2)d\sigma^2 \prod_{j=1}^p \frac{\lambda^2}{2}e^{-\lambda^2\tau_j^2/2}d\tau_j^2
\ee 
for $\sigma^2>0$ and $\tau_1^2,...,\tau_p^2>0$. 
\cite{Park:2008} suggested to use the improper prior $\pi(\sigma^2)\propto 1/\sigma^2$ to model the error variance.

As discussed in the introduction, the Lasso uses the same shrinkage for every coefficient and may not be consistent for certain design matrices in terms of
model selection. This motivates us to replace 
(\ref{eq3}) in the hierarchical structure by a more adaptive penalty
\be\label{eq3'}
\sigma^2,\tau_1^2,...,\tau_p^2\sim\pi(\sigma^2)d\sigma^2 \prod_{j=1}^p \frac{\lambda_j^2}{2}e^{-\lambda_j^2\tau_j^2/2}d\tau_j^2.
\ee 
The major difference of this formulation is to allow  different $\lambda^2_j$, one for each coefficient. Intuitively, if small penalty is applied to those
covariates that are important and large penalty is applied to those which 
are unimportant, the Lasso estimate, as the posterior mode, can be model selection consistent (Zou, 2006; Wang et al. 2007).
Indeed, as we will see in Section \ref{secAdaptiveShrinkage} and in later numerical experiments, in the posterior distribution,
the $\lambda_j$'s for zero $\beta_j$'s will be much larger than those $\lambda_j$'s for nonzero $\beta_j$'s.

The Gibbs sampling scheme follows Park and Casella (2008). For Bayesian 
inference, the full conditional distribution of $\beta$ is
multivariate normal with mean $A^{-1}X^Ty$ and variance $\sigma^2A^{-1}$, where
$A=X^TX+D_\tau^{-1}$. The full conditional for $\sigma^2$ is inverse-gamma with 
shape parameter $(n-1)/2+p/2$ and scale parameter $(y-X\beta)^T(y-X\beta)/2+\beta^TD_\tau^{-1}\beta/2$ and $\tau_1^2,...,\tau_p^2$ are conditionally independent, with
$1/\tau_j^2$ conditionally inverse-Gaussian with parameters
\[ \tilde{\mu}_j = \frac{\lambda_j \sigma}{|\beta_j|}~\text{and}~\tilde{\lambda}_j=\lambda_j^2
\]where the inverse-Gaussian density is given by
\[ f(x)=\sqrt{\tilde{\lambda}}{2\pi} x^{-3/2} \exp\left\{-\frac{\tilde{\lambda}(x-\tilde{\mu}_j)^2}{2(\tilde{\mu})^2x} \right\},~x>0.
\]
As observed in Park and Casella (2008), the Gibbs sampler with block updating
of $\beta$ and $(\tau_1^2,...,\tau_p^2)$ is very fast.

\subsection{Choosing the Bayesian Adaptive Lasso Parameters} 
We discuss two approaches for choosing BaLasso parameters in 
the Bayesian framework: the empirical Bayes (EB) method and the hierarchical Bayes (HB) 
approach using hyper-priors. The EB approach aims to estimate the $\lambda_j$
via marginal maximum likelihood, while the HB approach uses hyperpriors on
the $\lambda_j$ which enables posterior inference on these shrinkage parameters.

\paradot{Empirical Bayes (EB) Estimation}
A natural choice is to estimate the hyper-parameters $\lambda_j$ by marginal maximum likelihood.
However, in our framework, the marginal likelihood for the $\lambda_j$s is not available in closed form.
To deal with such a problem, 
\cite{Casella:2001} proposed a multi-step approach based on an EM algorithm with the expectation in the E-step
being approximated by 
the average from the Gibbs sampler. 
The updating rule then for $\lambda_j$ is easily seen to be
\be\label{eqEM}
\lambda_j^{(k)}=\sqrt{\dfrac{2}{E_{\lambda_j^{(k-1)}}(\tau_j^2|y)}}
\ee 
where $\lambda_j^{(k)}$ is the estimate of $\lambda_j$ at the $k$th stage and the 
expectation $E_{\lambda_j^{(k-1)}}(.)$ is approximated by the average from the Gibbs sampler
with the hyper-parameters are set to $\lambda_j^{(k-1)}$.

Casella's method may be computationally expensive because many Gibbs sampler runs are needed. 
\cite{Atchade:2009} proposed a single-step approach based on stochastic approximation which 
can obtain the MLE of the hyper-parameters using a single Gibbs sampler run.
In our framework, making the transformation $\lambda_j=e^{s_j}$, 
the updating rule for the hyper-parameters $s_j$ can be seen as (\citeauthor{Atchade:2009} 2009, Algorithm 3.1)
\beqn
s_j^{(n+1)}=s_j^{(n)}+a_n(2-e^{2s_j^{(n)}}\tau_{n+1,j}^2)
\eeqn    
where $s_j^{(n)}$ is the value of $s_j$ at the $n$th iteration,
$\tau_{n,j}^2$ is the $n$th Gibbs sample of $\tau_j^2$,
and $\{a_n\}$ is a sequence of step-sizes such that
\[a_n\searrow0,\;\;\sum a_n=\infty,\;\;\sum a_n^2<\infty.\]
In the following simulation, $a_n$ is set to $1/n$.
Strictly speaking, choosing a proper $a_n$ is an important problem of stochastic approximation
which is beyond the scope of this paper.
In practice, $a_n$ is often set after a few trials by justifying the convergence of iterations graphically.  

\paradot{Hierarchical Model}
Alternatively, $\lambda_j$s themselves can be treated as random variables and join the Gibbs updating by using an appropriate prior on $\lambda_j^2$. Here for 
simplicity and numerical tractability,
we take the following gamma prior \citep{Park:2008}
\be\label{lambdaprior}
 \pi(\lambda_j^2)=\frac{\delta^{r}}{\Gamma(r)}(\lambda_j^2)^{r-1} e^{-\delta\lambda_j^2}.
\ee
The advantage of using such a prior is that the Gibbs sampling algorithm can be easily implemented. More specifically, when this
prior is used, the full conditional of $\lambda_j^2$ is gamma with 
shape parameter $1+r$ and rate parameter $\tau_j^2+\delta$. This specification
allows $\lambda_j^2$ to join the other parameters in the Gibbs sampler. Although
the number of the penalty parameters $\lambda_j$ has increased to $p$ in BaLasso
from a single parameter in Lasso, the fact that the same prior is used
on these parameters greatly reduces the degrees of freedom in specifying the 
prior.

As a first choice, we can fix hyper-parameters $r$ and $\delta$ to some small values
in order to get a flat prior.
Alternatively, we can fix $r$ and use an empirical Bayes approach where $\delta$ is estimated.
The updating rule for $\delta$ (Casella, 2001) can be seen as
\beqn\label{EB1}
\delta^{(k)} = \frac{pr}{\sum_{j=1}^p E_{\delta^{(k-1)}}(\lambda_j^2|y)}.
\eeqn
Theoretically, we need not worry so much about how to select $r$
because parameters that are deeper in the hierarchy have less effect on inference \citep[p.260]{Lehmann:1998}. In our simulation study and data analysis, we use $r=.1$ which gives a fairly flat prior and stable results.

\subsection{Adaptive shrinkage}\label{secAdaptiveShrinkage}
By allowing different $\lambda_j^2$, adaptive shrinkage on the coefficients is possible.  
We demonstrate the adaptivity by a simple simulation 
in which a data set of size 50 is generated from the model
\[y=\beta_1x_1+\beta_2x_2+\sigma\epsilon\]
with $\beta=(3,\ 0)',\ \sigma=1,\ \epsilon\sim N(0,1)$.
\begin{figure*}
\centerline{\includegraphics[width=.8\textwidth]{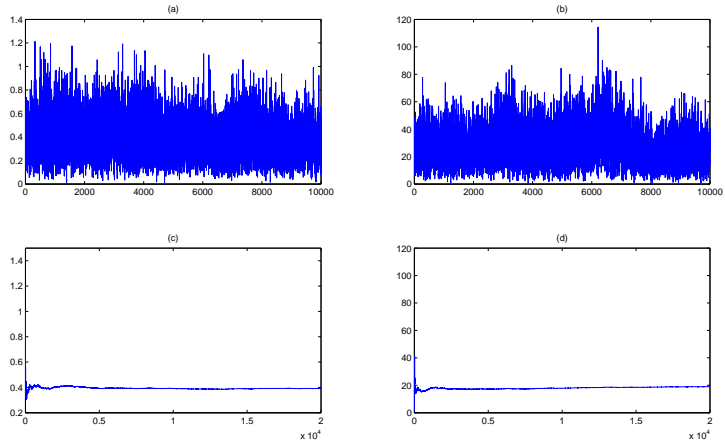}}
\caption{\label{Figure2}
(a)-(b): Gibbs samples for $\lambda_1$ and $\lambda_2$, respectively.
(c)-(d): Trace plot for $\lambda^{(n)}_1$ and $\lambda^{(n)}_2$ by Atchade's method.}
\end{figure*}

\begin{figure*}
\centerline{\includegraphics[width=.6\textwidth]{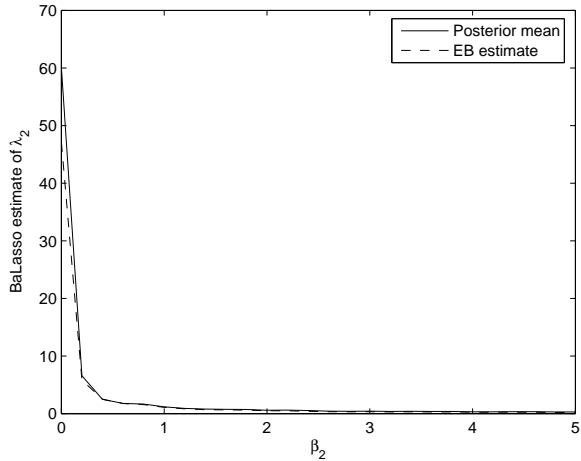}}
\caption{\label{Figure2beta}
Plots of EB and posterior estimates of $\lambda_2$ versus $\beta_2$}
\end{figure*}

Because $\beta_1\not=0,\ \beta_2=0$ we expect that the EB and posterior estimate of 
$\lambda_2$ would be much larger than that of $\lambda_1$. As a result, a heavier penalty is put on $\beta_2$ such that $\beta_2$ is more likely to be shrunken to zero. This phenomenon is demonstrated graphically in Figure \ref{Figure2}.
Figure \ref{Figure2} (a)-(b) plot 10,000 Gibbs samples (after discarding 10,000 {\it burn-in} samples) 
for $\lambda_1$ and $\lambda_2$ (not $\lambda_1^2,\ \lambda_2^2$), respectively.
The posterior distribution of $\lambda_2$ is central around a value of $22$
which is much larger than .39, the posterior median of $\lambda_1$. 
Figure \ref{Figure2} (c)-(d) shows the trace plots of iterations $\lambda_1^{(n)}$, $\lambda_2^{(n)}$
from Atchade's method. Marginal maximum likelihood estimates 
of $\lambda_1$ and $\lambda_2$ are 0.39 and 19, respectively. 
In Figure \ref{Figure2beta} we plot EB and posterior mean estimates of $\lambda_2$ versus $\beta_2$ when $\beta_2$ varies from 0 to 5. Clearly,
both the EB and the posterior estimates of $\lambda_2$ decrease as $\beta_2$ increases, which demonstrates that lighter penalty is applied for stronger signals.

\section{Inference}\label{secInf}
\subsection{Estimation and Model Selection}\label{ModelSelection}
For the adaptive Lasso, the usual methods to choose the $\lambda_j$'s would be
computationally demanding. 
From the Bayesian perspective, one can draw MCMC samples based on BaLasso
and get an estimated posterior quantity for $\beta$.
Like the original Bayesian Lasso, however, a full posterior exploration gives
no sparse models and would fail as a model selection method.
Here we take a hybrid Bayesian-frequentist point of view in which 
coefficient estimation and variable selection are simultaneously conducted 
by plugging in an estimate of $\lambda$ into \eqref{eq2.1}, where $\lambda$ might be the marginal maximum
likelihood estimator, posterior median or posterior mean.  
Hereafter these suggested strategies are abbreviated as BaLasso-EB, BaLasso-Median, and BaLasso-Mean, respectively.

With the presence of a posterior sample,  we also propose another strategy for exploring model uncertainty. Let $\{\lambda^{(s)}\}_{s=1}^N$ be Gibbs samples drawn from the hierarchical model \eqref{eq2}, \eqref{eq3'} and \eqref{lambdaprior}.
For the $s$th Gibbs sample $\lambda^{(s)}=(\lambda_1^{(s)},...,\lambda_p^{(s)} )'$, 
we plug $\lambda^{(s)}$ into (\ref{eq2.1}) and then 
record the frequencies of each variable being chosen out of $N$ samples. 
The final chosen model consists of those variables whose frequencies are not less than 0.5.
This strategy will be abbreviated as BaLasso-Freq.
The chosen model is somewhat similar in spirit to the so-called {\it median probability (MP) model} proposed by \cite{Barbieri:2004}.

As we will see in Section \ref{secExam}, all of our proposed strategies
have surprising improvement in terms of variable selection 
over the original Lasso and the adaptive Lasso.

\subsection{A Model Averaging Strategy}\label{secBMA}
When model uncertainty is present, making inferences based on a single model
may be dangerous. Using a set of models helps to account for this uncertainty and can provide improved inference. In the Bayesian framework, Bayesian model averaging (BMA) is widely used for prediction.  BMA generally provides
better predictive performance than a single chosen model,
see \cite{Raftery:1997,Hoe:99} and references therein. For making inference via multiple models, we use the hierarchical model approach for estimating $\lambda$ and refer to the strategy outlined below as BaLasso-BMA.
It should be emphasized, however, that our model averaging strategy is unrelated to the usual formal
Bayesian treatment of model uncertainty.  Rather, our idea is simply to use an ensemble of sparse models
for prediction obtained from sampling the posterior distribution of smoothing parameters and considering
different sparse conditional mode estimates of regression coefficients for the smoothing parameters so obtained. 

Let $\Delta=(x_\Delta,y_\Delta)$ be a future observation
and $D=(X,y)$ be the past data. 
The posterior predictive distribution of $\Delta$ is given by
\be\label{BMA}
p(\Delta|D)=\int p(\Delta|\beta)p(\beta|\lambda,D)d\beta p(\lambda|D)d\lambda.
\ee
Suppose that we measure predictive performance via a logarithmic scoring
rule \citep{Good:1952}, i.e., if $g(\Delta|D)$ is some distribution we use for prediction then our
predictive performance is measured by $\log g(\Delta|D)$ (where larger is better).
Then for any fixed smoothing parameter vector $\lambda_0$
\beqn
E(\log p(\Delta|D)-\log p(\Delta|\lambda_0,D))=\int \log\dfrac{p(\Delta|D)}{p(\Delta|\lambda_0,D)}p(\Delta|D)d\Delta
\eeqn
is nonnegative because the right hand side is the Kullback-Leibler divergence between $p(\Delta|D)$ and $p(\Delta|\lambda_0,D)$.  
Hence prediction with $p(\Delta|D)$ is superior in this sense to prediction 
with $p(\Delta|\lambda_0,D)$ with any choice of $\lambda_0$.

Our hierarchical model \eqref{eq2}, \eqref{eq3'} and \eqref{lambdaprior} offers a natural way 
to estimate the predictive distribution \eqref{BMA},
in which the integral is approximated by the average from Gibbs samples of $\lambda$.
For example, in the case of point prediction for $y_\Delta$ with squared error loss, 
the ideal prediction is
\[E(y_\Delta|D)=\int x_\Delta'E(\beta|\lambda,D)p(\lambda|D)d\lambda=x_\Delta'E(\beta|D),\]
where $E(\beta|D)$ can be estimated by the mean of Gibbs samples for $\beta$.
Write $\hat{\beta}_\lambda$ as the conditional posterior
mode for $\beta$ given $\lambda$.  
One could approximate $x_\Delta' E(\beta|D)$ by replacing $E(\beta|D)$ with the conditional posterior
mode $\hat{\beta}_{\hat{\lambda}}$ for some fixed value $\hat{\lambda}$ of $\lambda$.  
However, this ignores uncertainty in estimating the penalty parameters.  An alternative
strategy is to replace $E(\beta|D,\lambda)$ in the integral above with
$\hat{\beta}_\lambda$ and to integrate it out accordingly. 
This should provide a better approximation to the full Bayes solution than the approach
which uses a fixed $\hat{\lambda}$.  
In fact, we predict  $E(y_\Delta|D)$ by $s^{-1}\sum_{i=1}^s x_\Delta'\hat{\beta}_{\lambda^{(i)}}$ where
$\lambda^{(i)}$, $i=1,...,s$, denote MCMC samples drawn from the posterior distribution
of $\lambda$.  Note that this approach has advantages in interpretation 
over the fully Bayes' solution.  By considering the models selected by the conditional
posterior mode for different draws of $\lambda$ from $p(\lambda|y)$ we gain an ensemble
of sparse models that can be used for interpretation.  
As will be seen in Section \ref{secExam}, when there is model uncertainty,
BaLasso-BMA provides an ensemble of sparse models and
may have better predictive performance
than conditioning on a single fixed smoothing parameter vector $\lambda$.

\section{Examples}\label{secExam}
In this section we study the proposed methods through numerical examples.
These methods are also compared to Lasso, aLasso and BLasso in terms of variable selection and predictions.
We use the LARS algorithm of \cite{Efron:2004} for Lasso and aLasso
in which fivefold cross-validation is used to choose shrinkage parameters.
In the adaptive Lasso, we either use the least squares estimate (Example 1 and 2) or the Lasso estimate (Example 3) as the preliminary estimate.
For the optimization problem \eqref{eq2.1}, we use the gradient projection algorithm developed by \cite{Figueiredo:2007}.

\subsection{Simulation}
\paradot{Example 1 (Simple example)}
We simulate data sets from the model
\begin{equation}\label{modelI}
y=x'\beta+\sigma\epsilon,
\end{equation}
where $\beta=(3,\ 1.5,\ 0,\ 0,\ 2,\ 0,\ 0,\ 0)'$, $x_j$ follows N(0,1) marginally and 
the correlation between $x_j$ and $x_k$ is $0.5^{|j-k|}$, and $\epsilon$ is iid N(0,1).
We compare the performance of the proposed methods in Section \ref{ModelSelection} 
to that of the original Lasso and adaptive Lasso.
The performance is measured by the frequency of correctly-fitted models over 100 replications.
The simulation results are summarized in Table \ref{tab1} and suggest that
the proposed methods perform better than Lasso and aLasso in model selection.

\begin{table}
  \begin{center}
    \begin{tabular}{cccccccccc}
$n$	 &$\sigma$	&Lasso		&aLasso	&BaLasso-Freq	&BaLasso-Median&BaLasso-Mean&BaLasso-EB\\\hline
30	 &1		&50		&71		&86		&	86	  &	97&78\\
	 &3		&17		&8		&35		&	34	  &	18&39\\
60	 &1		&66		&76		&81		&	79	  &	100&83\\
	 &3		&44		&38		&54		&	53	  &	55&46\\
120	 &1		&73		&76		&87		&	87	  &	100&87\\
	 &3		&58		&55		&81		&	81	  &	97&86\\\hline
    \end{tabular}
  \end{center}
  \caption{Frequency of correctly-fitted models over 100 replications for Example 1.} \label{tab1}
\end{table}

\paradot{Example 2 (Difficult example)}
For the second example, we use Example 1 in  \cite{Zou:2006}, for which the 
Lasso does not give consistent model selection, regardless of the sample size and how the tuning parameter $\lambda$ is chosen. 
Here $\beta=(5.6,\ 5.6,\ 5.6,\ 0)'$ and the correlation
matrix of $x$ is such that $\text{cor}(x_j,x_k)=-.39,~j<k<4$ and $\text{cor}(x_j,x_4)=.23,~j<4$.

The experimental results are summarized in Table \ref{tab4}
in which the frequencies of correct selection are shown. 
We see that the original Lasso does not seem to give consistent model selection.
For all the other methods, the frequencies of correct selection go to 1 as $n$ increases and $\sigma$ decreases.
In general, our proposed method for model selection performs better than aLasso.  
 
\begin{table}
  \begin{center}
    \begin{tabular}{cccccccc}
$n$ & $\sigma$ & Lasso & aLasso & BaLasso-Freq & BaLasso-Median & BaLasso-Mean & BaLasso-EB\\\hline
60&9&0&5&8&8&9&12\\
120&5&10&45&66&65&66&51\\
300&3&12&65&83&83&85&83\\
300&1&12&100&100&100&100&100\\
\hline
    \end{tabular}
  \end{center}
  \caption{Frequency of correctly-fitted models over 100 replications for Example 2.} \label{tab4}
\end{table}

\paradot{Example 3 (Large $p$ example)}
The variable selection problem with large $p$ (even larger than $n$) is recently an active research area.
We consider an example of this kind in which $p=100$ with 
various sample sizes $n=50,\ 100,\ 200$.
We set up a {\em sparse recovery problem} in which most of coefficients are zero
except $\beta_{j}=5,\ j=10,20,...,100$.
From the previous examples, the performances of the four methods BaLasso-Freq, BaLasso-Median, BaLaso-Mean and BaLasso-EB are similar. We therefore just consider the BaLasso-Mean as a representative and compare it to the adaptive Lasso
which is generally superior to the Lasso.

Table \ref{tab4'} summarizes our simulation results,
in which the design matrix is simulated as in Example 1. 
BaLasso-Mean performs satisfactorily in this example and outperforms aLasso 
in variable selection.

\begin{table}
  \begin{center}
    \begin{tabular}{cccccccccc}
$n$	&$\sigma$	&aLasso		&BaLasso-Mean\\ 
\hline
50	&1		&24		&39\\
	&3		&24		&35\\
	&5		&8		&29\\
\hline
100	&1		&40		&100\\
	&3		&39		&99\\
	&5		&20		&86\\
\hline
200	&1		&100		&100\\
	&3		&88		&100\\
	&5		&78		&97\\
\hline
    \end{tabular}
  \end{center}
  \caption{Frequency of correctly-fitted models over 100 replications for Example 3.} \label{tab4'}
\end{table}

\paradot{Example 4 (Prediction)} 
In this example, we examine the predictive ability of BaLasso-BMA experimentally.
As discussed in Section \ref{secBMA},
when there is model uncertainty,
making predictions conditioning on a single fixed parameter vector is not optimal predictively.
Suppose that the dataset $D$ is split into two sets: a {\it training set} $D^T$ and {\it prediction set} $D^P$.
Let $\Delta=(x_\Delta,y_\Delta)\in D^P$ be a future observation and
$\hat y_\Delta$ be a prediction of $y_\Delta$ based on $D^T$.
We measure the predictive performance by the prediction squared error (PSE)
\be\label{eqpse}
\text{PSE}=\dfrac{1}{|D^P|}\sum_{\Delta\in D^P}|y_\Delta-\hat y_\Delta|^2.
\ee
We compare PSE of BaLasso-BMA to that of BaLasso-Mean in which 
$\hat y_\Delta=x_\Delta'\hat\beta$ where $\hat\beta$ is the solution to \eqref{eq2.1} 
with smoothing parameter vector fixed at the posterior mean of $\lambda$.
We also compare the predictive performance of BaLasso-BMA to that of the Lasso, aLasso,
and the original Bayesian Lasso (BLasso).
The implementation of BLasso is similar to BaLasso except
that BLasso has a single smoothing parameter. 

We first consider a small-$p$ case in which
data sets are generated from model \eqref{modelI} but now with $\beta=(3,\ 1.5,\ 0.1,\ 0.1,\ 2,\ 0,\ 0,\ 0)'$.
By adding two small effects we expect there to be model uncertainty.
Table \ref{table5} presents the prediction squared errors 
averaged over 100 replications with various factors $n_T$ (size of training set), 
$n_P$ (size of prediction set) and $\sigma$. 
The experiment shows that BaLasso-BMA performs slightly better than BLasso and BaLasso-Mean,
and much better than the Lasso and aLasso.

Similarly, we consider a large-$p$ case as in Example 3 but now with $\beta_{10}=\beta_{20}=\beta_{30}=\beta_{40}=\beta_{50}=.5$
in order to get model uncertainty. The results are summarized in Table \ref{table5'}.
Unlike for the small-$p$ case, BLasso now performs surprisingly badly. 
This may be due to the fact that BLasso uses the same shrinkage for every coefficient.
As shown, BaLasso-BMA outperforms the others.

 \begin{table}
  \begin{center}
    \begin{tabular}{ccccccc}
$n_T=n_P$&$\sigma$	&Lasso		&aLasso		&BLasso		&BaLasso-Mean	&BaLasso-BMA\\\hline
30	 &1		&2.029		&1.976		&1.276		&1.175		&1.165\\
	 &3		&17.43		&17.37		&10.88		&15.51		&11.06\\
	 &5		&42.74   	&42.13		&29.43		&41.32		&29.56\\
	 &10		&126.6  	&126.2		&109.6		&123.9		&109.9\\	
100	 &1		&1.449	    	&1.436		&1.044		&1.077		&1.032\\
	 &3		&12.69   	&12.58		&9.662		&9.627		&9.485\\
	 &5		&34.89   	&34.79		&25.79		&27.55		&25.83\\
	 &10		&117.6  	&117.5		&105.7		&118.2		&106.5\\
200	 &1		&1.279    	&1.274		&1.018		&1.036		&1.014\\
	 &3		&11.44	   	&11.40		&9.424		&9.326		&9.320\\
	 &5		&31.30   	&31.18		&25.32		&25.36		&25.19\\
	 &10		&120.7  	&120.7		&103.9		&108.8		&104.3\\
\hline 
   \end{tabular}
  \end{center}
  \caption{Prediction squared error averaged over 100 replications for the small-$p$ case. 
}\label{table5}
\end{table}

 \begin{table}
  \begin{center}
    \begin{tabular}{ccccccc}
$n_T=n_P$&$\sigma$	&Lasso		&aLasso		&BLasso		&BaLasso-Mean	&BaLasso-BMA\\\hline
100	 &1		&3.501    	&4.173		&9.574		&1.673		&1.234\\
	 &3		&15.49   	&17.70		&27.42   	&10.88		&10.42\\
	 &5		&34.45		&39.81		&42.43		&28.66		&28.19\\
	 &10		&149.3  	&178.1		&161.0		&124.5		&117.6\\	
200	 &1		&2.468    	&2.417		&5.231		&1.110		&1.072\\
	 &3		&17.11   	&17.09		&15.12  	&10.42 		&10.22\\
	 &5		&44.49   	&44.39		&33.92		&27.18		&27.06\\	
	 &10		&148.1  	&147.5		&136.1		&112.0		&108.9\\
\hline 
   \end{tabular}
  \end{center}
  \caption{Prediction squared error averaged over 100 replications for the large-$p$ case. 
}\label{table5'}
\end{table}

\subsection{Real Examples}

\paradot{Example 5: Body fat data}
Percentage of body fat is one important measure of health,
which can be accurately estimated by underwater weighing techniques.
These techniques often require special equipment and are sometimes not convenient,
thus fitting percent body fat to simple body measurements is a convenient way
to predict body fat. 
\cite{Johnson:1996} introduced a data set in which
percent body fat and 13 simple body measurements (such as weight, height and abdomen circumference) 
are recorded for 252 men
(see Table \ref{tabData} for the summarized data).  
This data set was also carefully analyzed by \cite{Hoe:99}. 
Following Hoeting et al., we omit the 42nd observation which is considered as an outlier. 
Previous diagnostic checking \citep{Hoe:99} showed that
it is reasonable to assume a linear regression model.

\begin{table}[ht]
\centering 
\begin{tabular}{clcc}
\hline
Predictor number & Predictor &mean&s.d. \\ 
\hline
$Y$&Percent body fat (\%)&18.89&7.72\\
$X_1$ &Age (years) &44.89 &12.63\\
$X_2$ &Weight (pounds) &178.82& 29.40\\
$X_3$ &Height (inches) &70.31& 2.61\\
$X_4$&Neck circumference (cm) &37.99 &2.43\\
$X_5$ &Chest circumference (cm) &100.80& 8.44\\
$X_6$&Abdomen circumference (cm)& 92.51& 10.78\\
$X_7$&Hip circumference (cm)& 99.84& 7.11\\\
$X_8$&Thigh circumference (cm) &59.36& 5.21\\
$X_9$&Knee circumference (cm) &38.57 &2.40\\
$X_{10}$&Ankle circumference (cm)& 23.10& 1.70\\
$X_{11}$ &Extended biceps circumference &32.27& 3.02\\
$X_{12}$&Forearm circumference (cm) &28.66& 2.02\\
$X_{13}$&Wrist circumference (cm) &18.23& .93\\
\hline
\end{tabular}
\caption{Body fat example: summarized data}
\label{tabData}
\end{table}

We first consider the variable selection problem. We center the variables so that the intercept is not considered. Lasso chooses $X_1,\ X_2,\ X_3,\ X_4,\ X_6,\ X_7,\ X_8,\ X_{11},\ X_{12},\ X_{13}$ in the final model with a BIC value $712.16$, while aLasso has one fewer variable $X_3$ with a BIC value $709.46$. BaLasso-Freq, BaLasso-Median, BaLasso-Mean and BaLasso-EB all choose $X_1,\ X_2,\ X_4,\ X_6,\ X_8,\ X_{11},\ X_{12},\ X_{13}$, one fewer variable ($X_7$) than aLasso. The BIC value for BaLasso is $708.92$, smaller than that of Lasso and aLasso.
A simple analysis shows that $X_3$ and $X_7$ are highly correlated to $X_6$
(the correlation coefficients are .89 and .92, respectively).
Additionally, $X_6$ is the most important predictor \citep{Hoe:99}.
Thus removing $X_3$ and $X_7$ from the model
helps to avoid the multicollinearity problem. To conclude, BaLasso chooses the
simplest model with the smallest BIC.

 \begin{table}[ht]
  \begin{center}
    \begin{tabular}{ccccccccccccc|c}
\hline
\multicolumn{13}{c|}{Models}&PMP (\%)\\
   $X_1$ &$X_2$&$X_3$ & $X_4$& $X_5$& $X_6$& $X_7$&$X_8$ &$X_9$ &$X_{10}$&$X_{11}$&$X_{12}$&$X_{13}$&\\
\hline
     1   &  1  &   0  &   1  &   0  &   1  &   0  &   1  &   0  &   0  &   1  &   1  &   1&    2.23\\
     1   &  1  &   0  &   0  &   0  &   1  &   0  &   1  &   0  &   0  &   0  &   1  &   1&    2.03\\
     1   &  1  &   0  &   0  &   0  &   1  &   0  &   0  &   0  &   0  &   1  &   0  &   1&    1.80\\
     0   &  1  &   0  &   0  &   0  &   1  &   0  &   0  &   0  &   0  &   1  &   0  &   1&    1.77\\
     1   &  1  &   0  &   1  &   0  &   1  &   0  &   1  &   0  &   0  &   0  &   1  &   1&    1.63\\
     1   &  1  &   0  &   1  &   0  &   1  &   0  &   0  &   0  &   0  &   1  &   0  &   1&    1.57\\
     1   &  1  &   0  &   1  &   0  &   1  &   1  &   1  &   0  &   0  &   1  &   1  &   1&    1.43\\
     0   &  1  &   0  &   1  &   0  &   1  &   0  &   0  &   0  &   0  &   1  &   0  &   1&    1.43\\
     0   &  1  &   0  &   0  &   0  &   1  &   0  &   0  &   0  &   0  &   0  &   1  &   1&    1.43\\
     0   &  1  &   0  &   0  &   0  &   1  &   0  &   1  &   0  &   0  &   0  &   1  &   1&    1.43\\
\hline
    \end{tabular}
  \end{center}
\caption{Body fat example: 10 models with highest posterior model probability
}\label{Ex1freq}
\end{table}

We now proceed to explore model uncertainty inherent in this dataset.
Let $M(\lambda)$ be the model selected w.r.t. shrinkage parameter vector $\lambda$.
We define the posterior model probability (PMP) of a model $M$ to be
\beqn
p(M|D)=\int_{\lambda:M(\lambda)=M}p(\lambda|D)d\lambda.
\eeqn
Note that this is not a posterior model probability in the usual sense in formal
Bayesian model comparison, but simply represents the uncertainty of the sparsity structure
in the conditional posterior mode estimate induced by the uncertainty in the posterior
distribution on the smoothing parameter.
From the Gibbs samples of $\lambda$, it is straightforward to estimate these PMPs.
Table \ref{Ex1freq} presents 10 models with highest PMP which indicates high model uncertainty.
The model with highest posterior probability and these 10 mostly selected models account for only 2.23\% and 16.8\% of the total posterior model probability, respectively.
With this model uncertainty, using a single model for prediction may be risky. 

We now examine the predictive performance of the approaches.
To this end, we split the dataset (without standardizing) into two parts:
the first 150 observations are used as the training set,
the remaining observations are used as the prediction set.
The out-of-sample predictive squared errors (PSEs)
of aLasso, BaLasso-Mean, BaLasso-Median, BaLasso-EB, BLasso and BaLasso-BMA are
18.92, 18.28, 19.79, 19.00, 18.69, 18.13, respectively. Thus, for this dataset,
BaLasso-BMA has the best predictive performance.

\paradot{Example 6: Prostate cancer data}
\cite{Stamey:89} studied the correlation between the level of
prostate antigen ({\it lpsa}) and a number of clinical measures in
men: log cancer volume ({\it lcavol}), log prostate weight ({\it lweight}), 
{\it age}, log of the amount of benign prostatic hyperplasia ({\it lbph}), 
seminal vesicle invasion ({\it svi}), log of capsular penetration ({\it lcp}), 
Gleason score ({\it gleason}), and percentage of Gleason scores 4 or 5 ({\it pgg45}). 
We assume a linear regression model between the response {\it lpsa} and the 8 covariates.
We first consider the variable selection problem.
The data set of size 97 is standardized so that the intercept $\beta_0$ is excluded.
Table \ref{Ex2model} summarizes the selected smoothing parameters and estimated coefficients
by various methods. Note that, for Lasso and aLasso there is just one smoothing parameter 
and putting the values on the first row as presented in the table
does not mean these parameters are only associated with the first predictor.
 \begin{table}[ht]
  \begin{center}
    \begin{tabular}{rrrrr|rrrrr}
\hline
\multicolumn{5}{c|}{Selected $\lambda$}&\multicolumn{5}{c}{Coefficient estimate $\hat\beta$}\\
BaLasso&BaLasso&BaLasso&Lasso	&aLasso	&BaLasso&BaLasso&BaLasso&Lasso	&aLasso\\
-EB	&-Median	&-Mean	&	&	&-EB	&-Median	&-Mean	&	&\\
\hline
1.24	& 1.19	&1.39	&2.40	&1.86	&0.563 & 0.562	&.563	&.561	&.568\\
1.59 	& 1.50	& 1.76	&	&	&0.436	& 0.436	&.436	&.357	&.437  \\
332.75	&841.05	& 1066	&	&	&  0	& 0	&0	&-.015	&0\\
55.78	&16.67	&20.41	&	&	&  0 	& 0	&0	&.1	&0   \\
1.15 	&1.08	&1.27	&	&	&0.587	& 0.594	&.580	&.432	&.510\\
97.61	&86.56	&113.2	&	&	&  0	& 0	&0	&0	&0\\
89.77	&78.69	&105.12	&	&	&  0	& 0	&0	&0	&0\\
754.38	&1241.70&1823.7	&	&	&  0	& 0	&0	&.005	&0\\
\hline
    \end{tabular}
  \end{center}
\caption{Prostate cancer example: selected smoothing parameters and coefficient estimates
}\label{Ex2model}
\end{table}

 \begin{table}[ht]
  \begin{center}
    \begin{tabular}{cccccccc|c}
\hline
\multicolumn{8}{c|}{Models}&PMP (\%)\\
\hline
    1	&	2	&		&		&	5	&		&		&		&	27.9\\
    1	&	2	&		&		&	5	&		&		&	8	&	16.1\\
    1	&		&		&	4	&	5	&		&		&		&	6.3\\
    1	&	2	&		&	4	&	5	&		&		&	8	&	5.9\\
    1	&	2	&		&		&		&		&		&	8	&	5.7\\
    1	&	2	&		&	4	&	5	&		&		&		&	5.1\\
    1	&	2	&	3	&		&	5	&		&		&	8	&	4.9\\
    1	&	2	&	3	&	4	&	5	&		&		&	8	&	4.9\\
    1	&		&		&	4	&	5	&		&		&	8	&	3.2\\
    1	&	2	&		&		&		&		&		&		&	3.1\\
  \hline
    \end{tabular}
  \end{center}
\caption{Prostate cancer example: 10 models with highest posterior model probability
}\label{Ex2freq}
\end{table}

The EB estimation here is implemented using the stabilized Algorithm 2.2 of \cite{Atchade:2009},
in which the compact sets are selected to be $\otimes [-n-1,n+1]$,
and the step-size $a_n=2/n$ is obtained after a few trials by justifying the convergence of iterations $\lambda^{(n)}$ graphically. As shown in Table \ref{Ex2model},
BaLasso-EB, BaLasso-Mean and BaLasso-Median give very similar estimates for $\lambda_j$ corresponding to nonzero coefficients,
but fairly different estimates for $\lambda_j$ corresponding to zero coefficients. The effects of increased penalty parameters on the zero coefficients are
obvious: smaller shrinkage is applied to the nonzero coefficients and
larger shrinkage is applied to those which should be removed.  

The adaptive Lasso and all of the proposed strategies (including BaLasso-Freq also)
for variable selection produce the same model whose BIC is -25.19,
while BIC of the model selected by Lasso is -21.38.
Therefore the model chosen by our methods is favorable.

Table \ref{Ex2freq} presents 10 models with highest PMP.
The mostly selected model is the same as the one selected by aLasso and our methods.
In comparison to the previous example, the presence of model uncertainty is not very clear in this case.
The model with highest posterior probability accounts for 27.9\% of the total
which is considerably large.
Moreover, this probability is also considerably different from that of the model with second highest posterior probability.

To examine the predictive performance,
we split the data set (without standardizing) into two sets: the first 50 observations form the training set $D^T$,
the rest form the prediction set $D^P$.
The PSEs of aLasso, BLasso, BaLasso-Median, BaLasso-BMA are 1.89, 1.91, 1.91, 1.86 respectively.
Therefore, although the presence of model uncertainty is not very clear, 
BaLasso-BMA still provides comparable and slightly better estimates in terms of prediction.

\section{A Unified Framework}\label{unified}
So far, we have focused on BaLasso for linear regression. This section
extends the BaLasso to more complex models such as generalized linear models, Cox's models and so on, with other penalties, such as the group penalty (Yuan and Lin, 2006) and the composite absolute penalty (Zhao, Rocha and Yu, 2009). This unified framework enables us to study variable selection in a much broader context.
  
 Denote by $L(\beta)$ the minus log-likelihood. In order to use the BaLasso developed for linear regression, we approximate $L(\beta)$ by the least squares approximation (LSA) in Wang and Leng (2007) 
\begin{eqnarray*}
L(\beta)&\approx& L(\tilde{\beta})+\frac{\partial L(\tilde{\beta})}{\partial\beta}(\beta-\tilde{\beta})+\frac12(\beta-\tilde{\beta})'\frac{\partial^2L(\tilde{\beta})}{\partial\beta^2}(\beta-\tilde{\beta})\\
&=&\text{constant}\ + \frac12 (\beta-\tilde{\beta})\hat\Sigma^{-1}(\beta-\tilde{\beta})
\end{eqnarray*}
where $\tilde{\beta}$ is the MLE of $\beta$ and $\hat\Sigma^{-1}:={\partial^2L(\tilde{\beta})}/{\partial\beta^2}$. To use the BaLasso for a general model, the sampling distribution of $y$, conditional on $\beta$, can be approximately written as
\[ y|\beta\sim\exp\left(-\frac12(\beta-\tilde{\beta})'\hat\Sigma^{-1}(\beta-\tilde{\beta})\right).
\]
And we only need to update the hierarchical model for $y$ in the linear model using this expression while keeping other specifications intact.
Now we discuss in detail three novel applications of BaLasso for models with flexible penalties.

\paradot{BaLasso with LSA}
The frequentist adaptive Lasso for general models estimates $\v\b$ by minimizing
\be\label{generalLasso}
L(\v\b)+\sum\l_j|\b_j|.
\ee
Its Bayesian version is the following
\bqan
\v y|\v\b &\sim& \exp\left(-\frac12(\v\b-\tilde{\v\b})'\hat\Sigma^{-1}(\v\b-\tilde{\v\b})\right),\\
\v\beta|\v\tau^2&\sim&N_p(\v 0,D_\tau), \; D_\tau=\mbox{diag}(\v\tau^2),\\
\v\tau^2|\v\l^2&\sim&\prod_{j=1}^p \frac{\l_j^2}{2}e^{-\lambda_j^2\tau_j^2/2},\\
\v\l^2&\sim&\prod_{j=1}^p(\lambda_j^2)^{r-1} e^{-\delta\lambda_j^2}
\eqan
where $\v\tau^2:=(\tau_1^2,...,\tau_p^2)',\ \v\l^2:=(\l_1^2,...,\l_p^2)'$.
Note that we no longer have $\s^2$ in the hierarchy. 
The full conditionals are specified by
\bqan
\v\b|\v y,\v\tau^2,\v\l^2 &\sim& N_p\left((\hat\S^{-1}+D_\tau^{-1})^{-1}\hat\S^{-1}\tilde{\v\b},(\hat\S^{-1}+D_\tau^{-1})^{-1}\right),\\
\frac{1}{\tau_j^2}=\g_j|\v y,\v\b,\v\l^2 &\sim& \text{inverse-Gaussian}\left(\frac{\l_j}{|\b_j|},\l_j^2\right),\ j=1,...,p,\\
\l_j^2|\v y,\v\b,\v\tau^2 &\sim& \text{gamma}(r+1,\delta+\frac{\tau_j^2}{2}),\ j=1,...,p.
\eqan

\paradot{BaLasso for group Lasso}
The adaptive group Lasso (Yuan and Lin, 2006) for general models minimizes
\be\label{generalgLasso}
L(\v\b)+\sum_{j=1}^J\l_j\|\v\b_j\|_{l_2}
\ee
where $\v\b_j$ is the coefficient vector of the $j$th group, $j=1,...,J$.
The corresponding Bayesian hierarchy is as follows: 
\bqan
\v y|\v\b &\sim& \exp\left(-\frac12(\v\b-\tilde{\v\b})'\hat\Sigma^{-1}(\v\b-\tilde{\v\b})\right),\\
\v\b_j|\v\tau^2&\sim&N_{m_j}(\v 0,\tau_j^2\I_{m_j}),\ j=1,...,J\\
\tau^2_j|\v\l^2&\sim&\text{gamma}\left(\frac{m_j+1}{2},\frac{\l_j^2}{2}\right),\ j=1,...,J\\
\l_j^2&\sim&\text{gamma}(r,\delta),\ j=1,...,J
\eqan
where $m_j$ is the size of group $j$, $\I_{m_j}$ is the identity matrix of order $m_j$. This prior was also used by Kyung et al. (2009) for grouped variable selection in linear regression.

The full conditionals can be obtained as follows. Let $\tilde X$ be the square root matrix of $\hat\S^{-1}$ and $\tilde{\v y}:=\tilde X\tilde{\v\b}$.
Write $\tilde X=[\tilde X_1,...,\tilde X_J]$ with block matrices $\tilde X_j$ of size $p\times m_j$. We have
\bqan
\v\b_j|\v y,\v\b_{-j},\v\tau^2,\v\l^2&\sim& N_{m_j}\left(A_j^{-1}\tilde X_j'(\tilde{\v y}-\sum_{j'\not=j}\tilde X_{j'}\v\b_{j'}),A_j^{-1}\right),\\
\frac{1}{\tau_j^2}=\g_j|\v y,\v\b,\v\l^2&\sim&\text{inverse Gaussian}\left(\frac{\l_j}{\|\v\b_j\|},\l_j^2\right),\\ 
\l_j^2|\v y,\v\b,\v\tau^2&\sim& \text{gamma}\left(r+\frac{m_j+1}{2},\delta+\frac{\tau_j^2}{2}\right),\ j=1,...,J,
\eqan 
where $\v\b_{-j}=(\v\b_1,...,\v\b_{j-1},\v\b_{j+1},...,\v\b_J)$ and $A_j=\tilde{X}_j'\tilde{X}_j+(1/\tau_j^2)\I_{m_j}$.

\paradot{BaLasso for composite absolute penalty}
We now consider the group selection problem in which
a natural ordering among the groups is present.
By $j\to j'$, we mean that group $j$ should be added into the model before another group $j'$,
i.e., if group $j'$ is selected then group $j$ must be included in the model as well. We extend the composite absolute penalty (Zhao, Rocha and Yu, 2009) by allowing different tuning parameters for different groups 
\[\sum_{\text{group}\ j}\l_j\|(\v\b_j,\v\b_{\text{all}\ j':j\to j'})\|_{l_2},\]
where $\beta_j$ is a coefficient vector and this penalty represents some hierarchical structure in the model.
From this, the desired prior for $\v\b$ is the multi-Laplace
\beqn
\pi(\v\b)\propto\exp\left(\sum_{j}\l_j\|(\v\b_j,\v\b_{j':j\to j'})\|_{l_2}\right)
\eeqn
which can be expressed as the following normal-gamma mixture
\be
\int\left(\dfrac{1}{2\pi\tau_j^2}\right)^{\frac{k_j}{2}}\exp\left(-\frac{\|(\v\b_j,\v\b_{j':j\to j'})\|^2}{2\tau_j^2}\right)\dfrac{(\frac{\l_j^2}{2})^{\frac{k_j+1}{2}}(\tau_j^2)^{\frac{k_j+1}{2}-1}}{\Gamma(\frac{k_j+1}{2})}\exp(-\dfrac{\l_j^2\tau_j^2}{2})d\tau_j^2=\exp\left({\l_j\|(\v\b_j,\v\b_{j':j\to j'})\|}\right) 
\ee
where $k_j:=m_j+\sum_{j':j\to j'}m_{j'}$.
Similar to the Bayesian formulations before,
this identity leads to the idea of using a hierarchical Bayesian formulation 
with a normal prior for $\v\b|\v\tau^2$ and a gamma prior for $\tau^2_j$.
More specifically, the prior for $\v\b|\v\tau^2$ will be
\beqn
\v\b|\v\tau^2\propto\exp\left(-\sum_{j}\dfrac{\|(\v\b_j,\v\b_{j':j\to j'})\|^2}{2\tau_j^2}\right)
=\prod_{j}\exp\left(-\frac12\Big(\frac1{\tau_j^2}+\sum_{j':j'\to j}\frac1{\tau_{j'}^2}\Big)\|\v\b_j\|^2\right).
\eeqn
This suggests that the hierarchical prior for $\v\b_j|\v\tau^2$ is independently normal
with mean $\v 0$ and covariance matrix $(1/\tau_j^2+\sum_{j':j'\to j}1/{\tau_{j'}^2})^{-1}\I_{m_j}$, $j=1,...,J$.
We therefore have the following hierarchy 
\bqan
\v y|\v\b &\sim& \exp\left(-\frac12(\v\b-\tilde{\v\b})'\hat\Sigma^{-1}(\v\b-\tilde{\v\b})\right),\\
\v\b_j|\v\tau^2&\sim&N_{m_j}\left(\v 0,\s_j^2\I_{m_j}\right),\;\;\text{where}\;\;\s_j^2:=(\frac1{\tau_j^2}+\sum_{j':j'\to j}\frac1{\tau_{j'}^2})^{-1}\\
\tau_j^2|\v\l^2&\sim&\text{gamma}\left(\frac{k_j+1}{2},\frac{\l_j^2}{2}\right)\\
\l_j^2&\sim&\text{gamma}(r,\delta)\;\;\text{for}\;\; j=1,...,J.
\eqan
\paradot{Full conditionals}
It is now straightforward to derive the full conditionals as follows
\bqan
\v\b_j|\v y,\v\b_{-j},\v\tau^2,\v\l^2&\sim& N_{m_j}\left(A_j^{-1}\tilde X_j'(\tilde{\v y}-\sum_{j'\not=j}\tilde X_{j'}\v\b_{j'}),A_j^{-1}\right),\\
\frac{1}{\tau_j^2}=\g_j|\v y,\v\b,\v\l^2&\sim&\text{inverse Gaussian}\left(\frac{\l_j}{\|(\v\b_j,\v\b_{j':j\to j'})\|},\l_j^2\right),\\ 
\l_j^2|\v y,\v\b,\v\tau^2&\sim& \text{gamma}\left(r+\frac{k_j+1}{2},\delta+\frac{\tau_j^2}{2}\right),\ j=1,...,J\\
\eqan
where $\v\b_{-j}=(\v\b_1,...,\v\b_{j-1},\v\b_{j+1},...,\v\b_J)$ and $A_j=\tilde{X}_j'\tilde{X}_j+(1/\s_j^2)\I_{m_j}$.

We now assess the usefulness of this unified framework by three examples. For brevity, we only report the performance of various methods in terms of model selection.        

\paradot{Example 7: BaLasso in logistic regression}
We simulate independent observations from Bernoulli distributions with probabilities of success
\[
\mu_i = P(y_i=1|x_i,\beta)=\frac{\exp(5+ x_i'\beta)}{1+\exp(5+x_i'\beta)}
\]
where $\beta=(3,\ 1.5,\ 0,\ 0,\ 2,\ 0,\ 0,\ 0)'$, and
$x_i=(x_{i1},...,x_{ip})\sim N_p(0,\Sigma)$ with $\sigma_{ij}=0.5^{|i-j|}$.
We compare the performance of the BaLasso to that of the Lasso and the aLasso. 
The performance is measured by the frequency of correct fitting and average number of zero
coefficients over 100 replications.
The weight vector in aLasso is as usual assigned as $\hat{w}=1/|\hat{\beta}^{(0)}|$, where $\hat{\beta}^{(0)}$ is the MLE.
The shrinkage parameters in Lasso and aLasso are tuned by 5-fold cross-validation.
Table \ref{tab1.0} presents the simulation result for various sample size $n$.
The aLasso in this example works better than the Lasso.
The suggested BaLasso works very well, especially when the sample size $n$ is large.  In addition, the BaLasso often produces sparser models than the others do. 
\begin{table}
  \begin{center}
    \begin{tabular}{cccc}
$n$	 	&Lasso		&aLasso		&BaLasso\\\hline
200	 	&3(2.15)	&35(3.97)	&36(6.19)		\\
300		&5(2.42)	&42(4.07)	&90(5.10)\\
500		&4(2.66)	&41(4.00)	&100(5.00)\\\hline
   \end{tabular}
  \end{center}
  \caption{Example 1: Frequency of correctly-fitted models over 100 replications. 
The numbers in parentheses are average numbers of zero-coefficients estimated. The oracle average number is 5.} \label{tab1.0}
\end{table}

\paradot{Example 8: BaLasso for group selection}
We consider in this example the group selection problem in a linear regression framework.
We follow the simulation setup of Yuan and Lin (2006).
A vector of 15 latent variables $\v Z\sim N_{15}(\v 0,\Sigma)$ with $\sigma_{ij}=0.5^{|i-j|}$
are first simulated.
For each latent variable $Z_i$, a 3-level factor $F_i$ is determined according to
whether $Z_i$ is smaller than $\Phi^{-1}(1/3)$, larger than $\Phi^{-1}(2/3)$ or in between.
The factor $F_i$ then is coded by two dummy variables.
There are totally 30 dummy variables $X_1,...,X_{30}$
and 15 groups with $\v\b_j=(\b_{2j-1},\b_{2j})',\ j=1,...,J=15$.
After having the design matrix $X$, a vector of responses is generated from the following linear model
\be\label{gLassomodel}
\v y = X\v\b+\v\epsilon,\;\;\v\epsilon\sim N_n(\v 0,\I)
\ee
where most of $\v\b_j=\v 0$ except $\v\b_1=(-1.2,\ 1.8)',\ \v\b_3=(1,\ 0.5)',\ \v\b_5=(1,\ 1)'$.
We compare the performance of the BaLasso to that of the gLasso in Yuan and Lin (2006) and the adaptive group Lasso (agLasso, Wang and Leng, 2008)
in terms of frequencies of correct fitting and average numbers of not-selected factors over 100 replications.
We follow Wang and Leng (2008) to take the weights $\hat w_j=1/\|\hat{\v\b}_j^\MLE\|$
with $\hat{\v\b}_j^\MLE$ are the MLE of $\v\b_j$.
The tuning parameters in gLasso and agLasso are tuned using AIC with the degrees of freedom as in Yuan and Lin (2006). We use
1000 values of $\l$ equally spaced from 0 to $\l_{\max}$ to search for the optimal value.
Table \ref{tab2.0} reports the simulation result.
Both gLasso and agLasso seem to select unnecessarily large models 
and have low rate of correct fitting.
In contrast, the BaLasso seems to produce more parsimonious models when $n$ is small.
In general, the BaLasso works much better than the others in terms of model selection consistency.

\begin{table}
  \begin{center}
    \begin{tabular}{cccc}
$n$	 	&gLasso		&agLasso		&BaLasso\\\hline
100	 	&5(6.64)	&22(9.60)	&15(14.86)		\\
200	 	&8(6.92)	&48(10.72)	&90(12.04)		\\
500		&7(7.24)	&70(11.34)	&100(12.00)\\\hline
   \end{tabular}
  \end{center}
  \caption{Example 8: Frequency of correctly-fitted models and average numbers (in parentheses) of not-selected factors over 100 replications. The oracle average number is 12. 
} \label{tab2.0}
\end{table}

\paradot{Example 9: BaLasso for main and interaction effect selection}
In this example we demonstrate the BaLasso with composite absolute penalty for selecting main and interaction effects in a linear framework.
We consider the model II of Yuan and Lin (2006).
First, 4 factors are created as in the previous example, each factor is then coded by two dummy variables.
The true model is generated from \eqref{gLassomodel} with main effects $\b_1=(3,\ 2)'$, $\b_2=(3,\ 2)'$ and interaction $\b_{1\cdot 2}=(1,\ 1.5,\ 2,\ 2.5)'$. There are totally 10 groups (4 main effects and 6 second-order interaction effects) with the natural ordering in which main effects should be selected before their corresponding interaction effects. 
We use the BaLasso formulation with composite absolute penalty to account for this ordering.
Table \ref{tab3.0} reports the simulation results.
We observe that both gLasso and agLasso sometimes select effects in a ``wrong" order (interactions are seclected while the corresponding main effects are not). As a result, they have low rates of correct fitting.
The BaLasso always produce the models with effects in the ``right" order.
This fact has been theoretically proven in Zhao, Rocha and Yu (2009).
In general, the BaLasso outperforms its competitors.
\begin{table}
  \begin{center}
    \begin{tabular}{cccc}
$n$	 	&gLasso		&agLasso		&BaLasso\\\hline
100	 	&18(4.25)	&45(5.45)	&72(7.28)		\\
200	 	&36(5.16)	&88(6.78)	&100(7.00)		\\
500		&34(5.24)	&96(6.92)	&100(7.00)\\\hline
   \end{tabular}
  \end{center}
  \caption{Example 9: Frequency of correctly-fitted models and average numbers (in parentheses) of not-selected effects over 100 replications. The oracle average number is 7. 
} \label{tab3.0}
\end{table}

\section{Conclusion}\label{secCon}
We have proposed the Bayesian adaptive Lasso which is novel in two aspects. First, we use an adaptive penalty and have proposed methods for tuning parameter selection and estimation.  Second, we have proposed to use the posterior mode of the regression coefficients given the shrinkage parameters from their posterior for model averaging. Our approach retains the attractiveness of the usual Lasso in producing sparse models, and that of the aLasso in giving consistent models. Moreover, due to its Bayesian nature, an ensemble of sparse models, produced as the posterior modes estimates, can be used for model averaging. Thus, our approach provides a novel and natural treatment of exploration of model uncertainty and predictive inference. Finally, we have proposed a unified framework which can be applied to select groups of variables (Yuan and Lin, 2006) and other constrained penalties (Zhao, Rocha and Yu, 2009) in more general models. Empirically, we have shown its attractiveness compared to its competitors.
The software implementing our method is freely available from the authors' homepage.

\renewcommand{\baselinestretch}{1}
\normalsize
\bibliographystyle{apalike}

\end{document}